\documentclass[%
 preprint,
 superscriptaddress,
 groupedaddress,
  amsmath,amssymb,
  aps,
 prl,
 ]{revtex4-1}
\usepackage{graphicx}
\usepackage{dcolumn}
\usepackage{bm}
\usepackage{hyperref}

\usepackage{mathptmx, mathrsfs}
\usepackage{enumitem}
\usepackage{verbatim}
\usepackage{amsmath}
\usepackage{xcolor}
\usepackage{mathtools}
\newcommand{\bra}[1]{\ensuremath{\left\langle#1\right|}}
\newcommand{\ket}[1]{\ensuremath{\left|#1\right\rangle}}

\begin{document}

\title{A Protocol for Spectroscopists to Isolate The Effect of Berry Geometric Magnetic Forces on Molecular Dynamics}
 \author{Zeyu Zhou}
 \affiliation{Department of Chemistry, University of Pennsylvania, Philadelphia, Pennsylvania 19104, U.S.A.}

 \author{Joseph E. Subotnik}
 \affiliation{Department of Chemistry, University of Pennsylvania, Philadelphia, Pennsylvania 19104, U.S.A.}

 \author{Hsing-Ta Chen}
 \email{hsingc@sas.upenn.edu}
 \affiliation{Department of Chemistry, University of Pennsylvania, Philadelphia, Pennsylvania 19104, U.S.A.}

%


\begin{abstract}
We propose a novel means to isolate and quantify the effects of Berry force on molecular dynamics using two reasonably strong continuous wave (CW) laser fields with frequencies $\omega$ and $2\omega$.
For molecules or materials with three frequency-matching bright transitions ($\ket{0}\rightarrow\ket{1}$, $\ket{1}\rightarrow\ket{2}$, $\ket{0}\rightarrow\ket{2}$) at frequencies ($\omega$, $\omega$, $2\omega$) respectively, the effects of Berry curvature can be isolated by varying the phase between the two laser fields ($\Delta \phi$) and monitoring the dynamics. 
Moreover, we find that the resulting chemical dynamics can depend critically on the sign of $\Delta \phi$; in other words, the effects of Berry curvature can be enormous.
Thus, this manuscript represents an unusual step forward towards using light-matter interactions to affect chemical dynamics, suggesting that topological concepts usually invoked in adiabatic quantum optics and condensed matter can be directly applied to non-adiabatic chemical excited state dynamics.
\end{abstract}

 \maketitle

\maketitle

\section{Introduction\label{sec:Introduction}}
Within the realm of solid-state electronic structure theory, it is now well-appreciated that how Bloch orbitals change their phase as a function of crystal momentum (i.e. Berry phase\cite{berry1984quantal}) is fundamentally tied to the geometry and topology of a given material and has direct consequences as far as experimental observables, such as the anomalous Hall effect\cite{haldane_berry_2004,zhang_experimental_2005,xiao_berry_2010,gritsev_dynamical_2012}. 
Furthermore, in the world of electronic transport, it is well understood that when a current runs through a molecule or through a material, geometric magnetic forces (which follows from Berry phase) can emerge leading to current-induced spin-orbit torques\cite{manchon_current-induced_2019,brataas_current-induced_2012} or runaway vibrational motion\cite{Lu2010_blowing}.
However, in the context of molecular systems far from a metal surface, the experimental consequences of Berry phase remain murkier. 
More precisely, as far as molecular electronic structure is concerned, one considers Berry phase only around a conical intersection (CI) where the integral of the phase change around the CI is non-zero (and a multiple of $\pi$) which is known as molecular Aharonov-Bohm physics\cite{yarkony_determining_1998,yarkony_diabolical_1996,xie_nonadiabatic_2016,guo_accurate_2016,xie_constructive_2017,yuan_observation_2018,xie2019up,xie_quantum_2020}; such interference effects have now been measured for some realistic molecules\cite{xie_nonadiabatic_2016,xie2019up}. 
And yet, for many systems, numerical investigations of non-adiabatic dynamics suggest that semiclassical simulations (e.g. Tully's surface hopping method\cite{tully1990molecular}) can recover many observables after wavepackets approach a CI \emph{even though these methods do not include geometric phase} \cite{ryabinkin_geometric_2013,ryabinkin_geometric_2017,ryabinkin_analysis_2014}.
Moreover, recent analysis based on exact factorization\cite{abedi2010exact,agostini2015exact} has argued that the overall effect of a molecular Berry phase disappears under a gauge transformation\cite{juanes-marcos_theoretical_2005,min_is_2014,agostini2018exact,curchod2017dynamics}, except in the limit of non-zero circular nuclear currents\cite{requist_molecular_2016}.
In the end, for theoretical chemists who think about molecular motion (rather than condensed matter physicists who think about periodic electronic structure), the experimental importance or relevance of molecular Berry phase still remains unclear. 

Now one key assumption is usually made in the context of molecular non-adiabatic dynamics: almost always we assume that the electronic Hamiltonian is \emph{real}-valued.
However, when the electronic Hamiltonian is \emph{complex}-valued, it is known that a non-zero Berry phase yields a Lorentz-like magnetic force arising from the imaginary part of the derivative coupling ($\vec{d}$) vector\cite{berry_chaotic_1993,takatsuka_exploring_2011,takatsuka_lorentz-like_2017}. 
This so-called geometric magnetic force acts as the first order correction to the Born-Oppenheimer approximation and will affect nuclear motion whenever the electronic Hamiltonian is not real-valued\cite{mead1979noncrossing,mead1979determination}, e.g. for molecular systems with complex spin-orbit coupling\cite{matsika_effects_2001,matsika_effects_2001-1,matsika_spin-orbit_2002}. 
Nevertheless, even though derivative couplings ($\vec{d}$) are often large (e.g. whenever the energy gap between levels becomes small), most chemists have always assumed that they can ignore such a magnetic force even in the context of spin-dependent nuclear phenomena (e.g. spin-vibronic intersystem crossing)\cite{penfold_spin-vibronic_2018, talotta2020internal}.
Traditional surface hopping does not account for a Berry magnetic force directly or indirectly\cite{tully1990molecular}.

\begin{figure}[t]
    \centering
    \includegraphics{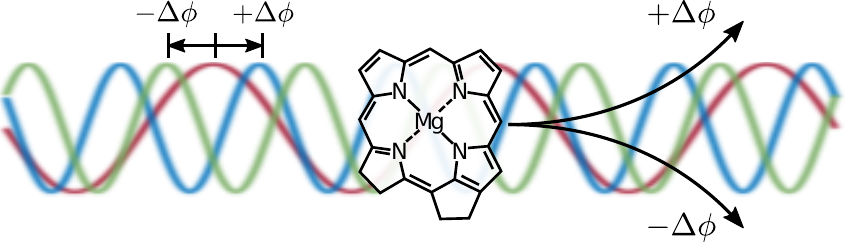}
    \caption{Schematic diagram of a molecular system (e.g. bacteriochlorophyll) under illumination by laser fields at frequencies $\omega$ (red) and $2\omega$ with phase difference $+\Delta\phi$ (blue) and $-\Delta\phi$ (green). Any detected difference in the resulting molecular dynamics can be attributed to Berry magnetic force exclusively. }
    \label{fig:schematic}
\end{figure}
In a future set of articles, we will suggest a set of \emph{complex}-valued Hamiltonians and chemical reactions for illustrating how Berry magnetic force can lead to spin-dependent nuclear motion.  
For the present letter, however, our goal is to demonstrate that Berry magnetic force should also be experimentally observable for a spin-less system with a \emph{real}-valued electronic Hamiltonian, provided that we have a strong laser source.
In order to generate a necessarily complex-valued Hamiltonian from a real-valued Hamiltonian, we will assume that the spin-less molecular system is coupled to multiple laser fields that are periodic in time with frequencies $\omega$ and $2\omega$ (see Fig.~\ref{fig:schematic}); such a system should be easily realizable with frequency doubling.
In such a case, using Floquet theory\cite{shirley_solution_1965}, one can expand the electronic wavefunction in a Floquet state basis (the electronic states dressed by $e^{im\omega t}$ for an integer $m$) and recast the explicitly time-dependent real-valued Hamiltonian ($\hat{H}(t)$) into a time-independent complex-valued Floquet Hamiltonian ($\hat{H}_F$).
Furthermore, the complex-valued nature of the Floquet Hamiltonian can be tuned by the phase difference between the two frequency components $\omega$ and $2\omega$ and, as we will show, this phase difference gives experimentalists the capacity to measure the dynamics of $\hat{H}_F$ and $\hat{H}_F^*$ independently.
By comparing the dynamics of $\hat{H}_F$ and $\hat{H}_F^*$, one can definitively isolate the effect of Berry phase on a laser-driven molecular system. 

Now using a laser-driven periodic potential (together with Floquet theory) to engineer materials to achieve a given band structure with desirable Berry phase is hardly new in the context of designing band structure\cite{flaschner2016experimental,oka2019floquet}; moreover, observing the
presence of a non-trivial Berry phase is considered a fingerprint of a light-induced conical intersection (LICI) in photodissociation dynamics\cite{halasz2011conical,halasz2012light,halasz2015direct,bouakline2018unambiguous,kim2012control}.
That being said, to our knowledge, the experimental ramifications of Berry force per se  have not been explored in the literature in the context of laser-driven nuclear dynamics.
For the most part, heretofore semiclassical simulations of laser-driven dynamics have either centered around monochromatic periodic potentials (for which the Floquet Hamiltonian is real-valued and there is no Berry geometric force\cite{kim2015ab,makhov2018floquet,maitra2002floquet,fiedlschuster_floquet_2016,fiedlschuster_surface_2017,zhou_nonadiabatic_2020,chen2020proper,restrepo_driven_2016}), or focused on electronic symmetry breaking of charge transport driven by an $\omega+2\omega$ field (where Berry force is not a key factor in these calculations)\cite{franco2006laser,franco2008femtosecond,franco2008laser}.
The goal of this letter is to establish a clear set of guidelines outlining (1) how to construct a Hamiltonian (or search for a molecular system) for which Berry force effects should be strong, and (2) how to experimentally observe Berry force using a continuous wave laser source.


\section{Floquet theory and Berry curvature}
Let us briefly review Floquet theory as applied to solving periodically driven nuclear-electronic system. Consider a real-valued electronic Hamiltonian that is periodic in time i.e. $\hat{H}(t)=\hat{H}(t+T_0)$ in a diabatic electronic basis $\ket{j}$ for $j=1,\cdots,N$. The corresponding time-independent Floquet Hamiltonian can be determined by taking the Fourier transform of $\hat{H}_F=\hat{H}-i\hbar\frac{\partial}{\partial t}$:
\begin{equation}
\bra{k n}\hat{H}_{F}\ket{j m}
=\frac{1}{T_{0}}\int_{0}^{T_{0}}dt\bra{k}\hat{H}(t)\ket{j}e^{-i(n-m)\omega t} + \delta_{jk}\delta_{m n}n\hbar\omega
\label{eq:floquettransform}
\end{equation}
Here $\omega = 2\pi/T_{0}$ and $\ket{j m}=\ket{j}e^{im\omega t}$ is an element of the diabatic Floquet state basis. 
Without loss of generality, we consider a time-periodic Hamiltonian written in a diabatic electronic basis of the form $H(t)=H_0+2\sum_{k=1}^\infty{H}_k\cos(k\omega t+\phi_k)$ where $H_k$ is a real-valued $N\times N$ block matrix. 
The corresponding Floquet Hamiltonian matrix takes the form of
\begin{equation}\label{eq:H_F}
H_F = \left(
\begin{array}{c c c c c}
\ddots & & & & \\
 & H_0+\hbar\omega & H_1e^{i\phi_1} & H_2e^{i\phi_2}  & \\
 & H_1^{\dag}e^{-i\phi_1}  & H_0 & H_1e^{i\phi_1}  & \\
 & H_2^{\dag}e^{-i\phi_2}  & H_1^{\dag}e^{-i\phi_1}  & H_0-\hbar\omega & \\
 & & & & \ddots \\
\end{array} \right)
\end{equation}
In the Floquet diabatic representation, the electronic wavefunction can be written as $\ket{\Psi_F(t)}=\sum_{j}\sum_{m=-\infty}^{\infty}C_{j m}(t)\ket{j m}$, and the Schr\"{o}dinger equation becomes $i\hbar\frac{\partial}{\partial t}C = {H}_FC$ with a complex-valued Floquet Hamiltonian ${H}_F$. 
Note that here the Floquet expansion dresses the electronic diabatic states and does not affect the nuclear degrees of freedom.

Next, in order to account for nuclear motion, we let $\hat{H}_F=\hat{H}_F(\vec{R})$ depend on a nuclear coordinate $\vec{R}$.
The Floquet Hamiltonian can be diagonalized by solving $\hat{H}_F(\vec{R})|\Phi^{J}(\vec{R})\rangle =V_F^{J}(\vec{R})|\Phi^{J}(\vec{R})\rangle$. 
Here $V_F^{J}(\vec{R})$ denotes the Floquet quasi-energy surface and $|\Phi^{J}(\vec{R})\rangle=\sum_{j}\sum_{m=-\infty}^{\infty}G^{J}_{j m}(\vec{R})\ket{j m}$ are the Floquet adiabatic states. 
Then, within the Floquet adiabatic representation, we can propagate non-adiabatic nuclear dynamics where the potential energy surfaces are $V_F^{J}(\vec{R})$ and the derivative coupling between Floquet adiabatic states $J$ and $K$ is
\begin{equation}
    \vec{d}_{JK}=\sum_{j}\sum_{m=-\infty}^{\infty}G^{J*}_{j m}\frac{\partial}{\partial\vec{R}}G^{K}_{j m} = \frac{\bra{G^{J}}\nabla_{\vec{R}} H_{F}\ket{G^{K}}}{\epsilon_{K} - \epsilon_{J}}
    \label{eq:derivcoup}
\end{equation}

With the complex-valued Floquet Hamiltonian, it is well known that the Berry curvature near the derivative coupling region yields an effective "magnetic" force on adiabatic surface $J$\cite{sakurai2014modern, shankar2012principles, miao2019extension} $\vec{F}_{J}^{mag} = \hbar\frac{\vec{P}}{M}\times\vec{B}_{J}$
where $\vec{P}$ is the nuclear momentum and $M$ is the nuclear mass. 
Here $\vec{B}_{J}$ is defined as the Berry curvature of Floquet adiabat $J$
\begin{equation}
    \vec{B}_{J} = -i\sum_{K\neq J}\vec{d}_{JK}\times \vec{d}_{KJ}.
    \label{eq:berrycurvature}
\end{equation}
Note that $\vec{d}_{JK}=-\vec{d}_{KJ}^*$ and the "magnetic" force arise from the imaginary part of the derivative coupling;
explicitly 
\begin{equation}
    \vec{F}_{J}^{mag} =\hbar\frac{\vec{P}}{M}\times\vec{B}_{J}= 2\hbar\text{Im}\sum_{K\neq J}\vec{d}_{JK}(\frac{\vec{P}}{M}\cdot\vec{d}_{KJ}).
\end{equation}

\section{Sufficient condition for observing Berry force}\label{sec:condition}
For a spectroscopist who wants to observe Berry curvature directly through nuclear dynamics, we will now establish a practical framework.
To begin this analysis, note that the Berry curvature of the Floquet Hamiltonian remains unchanged under two transformations: 
(1) adding an arbitrary phase factor to a spatial electronic basis function: $\ket{j}\rightarrow\ket{j}e^{i\theta_j}$ for $\theta_{j}\in[0,2\pi)$; 
(2) translating a Floquet diabatic basis function in time for arbitrary $\eta$: $e^{im\omega t}\rightarrow e^{im\omega (t+\eta/\omega)}$ .
If these operations are applied, a block of the Floquet Hamiltonian in \eqref{eq:H_F} transforms as follows 
\begin{equation}
    H_k\rightarrow\tilde{H}_k = U^\dagger H_k Ue^{ik\eta}
    \label{eq:shifttransform}
\end{equation}
Such a transformation (replacing $H_{k}$ with $\tilde{H_{k}}$) has no effect on the Berry curvature in Eqs.~\ref{eq:derivcoup} and \ref{eq:berrycurvature}, which will now allow us to isolate a set of sufficient conditions for isolating a Berry force.


\subsection*{Case \#1: Monochromatic Laser}
The simplest case to consider is the case of a monochromatic laser of the form $H(t)=H_0+H_1\cos(\omega t+\phi_1)$\cite{bajo2012mixed}.
In such a case, the Floquet Hamiltonian will be block tridiagonal (i.e. $H_1\neq0$, but $H_{2}=H_{3}=H_{4}=\dots=0$ in \eqref{eq:H_F}). 
If we now apply a transform of the form of \eqref{eq:shifttransform}, letting $\theta_j=0$ for all $j$ and $\eta=-\phi_1$, and noting that $\cos(\omega t)$ yields \emph{real} Fourier transform components ($\frac{1}{2}H_1$ and $\frac{1}{2}H_1$), we find that a complex-valued $H_F$ can always be transformed to a real-valued $\tilde{H}_F$. In other words, if a molecule and material is illuminated by a monochromatic laser, the Berry curvature is still strictly zero. 


\subsection*{Case \#2: 2 Frequencies, 2 Electronic States}
Next, we focus on the case where the laser profile includes two periodic frequency components. 
Specifically, we consider two electronic states ($N=2$) coupled through the laser excitations only (i.e. $H_0$ is diagonal). 
We find that $H_{F}$ cannot be transformed to a real-valued $\tilde{H}_F$ (except for a trivial case when $\phi_{2}-2\phi_{1}=0\mod\pi$), suggesting that one should be able to observe non-zero Berry force effect in this case.
However, if a system contains only two electronic states, there is only one energy difference that can be made resonant with one incoming frequency (and we have already discussed why Berry curvature is zero in the case of a monochromatic laser.) Thus, observing non-zero Berry curvature in the presence of only two electronic states would require higher order non-resonant light-matter interactions, for example, multiphoton absorption. Such effects are usually very weak and very likely will not yield robust experimental signals. 



\subsection*{Case \#3: 2 Frequencies, 3 Electronic States}
To observe a significant Berry force effect, we find we require a minimal model of three electronic states ($\ket{0}, \ket{1}, \ket{2}$) with three bright transitions ($\ket{0}\rightarrow\ket{1}$, $\ket{1}\rightarrow\ket{2}$ and $\ket{0}\rightarrow\ket{2}$). 
In this case, the complex-valued Floquet Hamiltonian cannot be made real-valued (except for the trivial case when $\phi_{2}-2\phi_{1}=0~\text{mod}~\pi$), such that a non-zero Berry curvature is expected.
Note that all three electronic transitions must be bright: if, for instance, the $\ket{1}\rightarrow\ket{2}$ transition is dark, (i.e. $(H_k)_{12}=0$ in \eqref{eq:H_F}), the Berry curvature will again likely be small as in case \#2 above. After all, with only two bright transitions and two frequencies, one can make a reasonable rotating wave approximation (RWA) on resonance and reduce the Floquet Hamiltonian to be effectively real-valued.
Thus, we are led to a Hamiltonian of the following mathematical form:
\begin{equation}\label{eq:tdhamiltonian}
    H(t) = H_0 + H_1 \cos\omega t + H_2 \cos(2\omega t+ \Delta\phi).
\end{equation}
where $\Delta\phi\in[-\pi,\pi]$. 

For the spectroscopist, control over the phase difference $\Delta\phi$ will allow one to isolate Berry curvature. 
When $\Delta\phi=0$ or $\pm\pi$, based on the above analysis, the Berry curvature is $0$. 
However, if $\alpha$ is a constant not equal to $0$ or $\pm\pi$, choosing $\Delta\phi=\alpha$ vs $\Delta\phi=-\alpha$ is equivalent to choosing $\hat{H}_{F}$ vs $\hat{H}^{*}_{F}$.
Most importantly, because two linear operators that are complex conjugate to each other, i.e. $\hat{H}_{F}$ vs $\hat{H}^{*}_{F}$, will have identical eigenvalues, any difference in the resulting nuclear dynamics must reflect differences in the phases of the eigenvectors, i.e. Berry force.
Thus, by comparing the dynamics of $\hat{H}_F$ and $\hat{H}_F^*$, the effect of Berry phase on a laser-driven molecular system can be isolated.

\section{Model and results}

For a numerical demonstration of a Berry force on nuclear dynamics using the Hamiltonian in \eqref{eq:tdhamiltonian}, we consider the following model with three electronic diabatic states $|j=0,1,2\rangle$ coupled to two nuclear degrees of freedom $\vec{R}=(x,y)$. 
$H_0$ is taken to be diagonal and of the form (see \eqref{eq:tdhamiltonian})
\begin{align}
\bra{0}H_0(\vec{R})\ket{0} &= A\tanh(x),\nonumber \\
\label{eq:T1st}
\bra{1}H_0(\vec{R})\ket{1} &= \hbar\omega-A\tanh(x), \\\nonumber
\bra{2}H_0(\vec{R})\ket{2} &= 0.2A + 2\hbar\omega 
\end{align}
with $A = 0.02$ and $\hbar\omega = 0.5$. 
We assume that the only coupling between the diabatic electronic states is caused by the coupling between the transition dipole moments and the external electric fields, which we take to be of the form  
\begin{equation}
\bra{j}H_1(\vec{R})\ket{k} =\bra{j}H_2(\vec{R})\ket{k} =  De^{-(x^2+y^2)/2\sigma^2}.
\label{eq:diabaticcoupling}
\end{equation}
for all $j\neq k$ with $\sigma = 1.0$ and $D=0.01$.

\begin{figure}[t]
    \centering
    \includegraphics{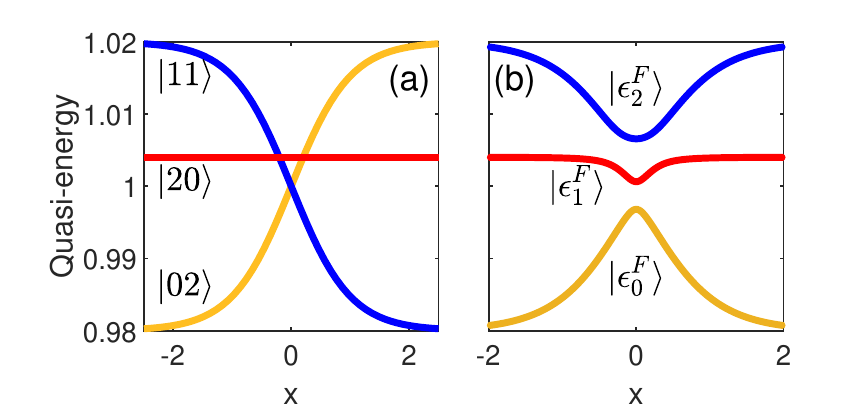}
    \caption{(a) The three diabatic Floquet states $\ket{02}$ (electronic state $\ket{0}$ dressed with $2$ photons), $\ket{11}$ (electronic state $\ket{1}$ dressed with $1$ photon) and $\ket{20}$ (electronic state $\ket{2}$ dressed with $0$ photons) approach each other near $x=0$. All other Floquet states have quasi-energies greater than 1.498 or smaller than 0.502 and are not dynamically relevant. (b) The energies of the three Floquet adiabatic states $\ket{\epsilon_{0}^{F}}$, $\ket{\epsilon_{1}^{F}}$ and $\ket{\epsilon_{2}^{F}}$, clearly show a complicated avoided crossing. All energies are evaluated at $y = 0$.}
    \label{fig: dia_ad_states}
\end{figure}

For convergence only, the eigenenergies of the Hamiltonian $H_0(\vec{R})$ have been chosen to mimic scattering potentials. Although all the theory above in Sec.~\ref{sec:condition} is completely general, working with a scattering Hamiltonian (where the diabatic Floquet states $|j m\rangle$ have constant asymptotic energies) will allow for a very simple visualization of the Floquet dynamics below. Moreover, for this Hamiltonian, invoking the RWA (where we only keep the three diabatic states $\ket{j m} \in \{\ket{02}$, $\ket{11}, \ket{20}\}$) is a fairly good approximation. For a visualization of the relevant diabatic and adiabatic energies, see Fig. \ref{fig: dia_ad_states}. Henceforward, the Floquet adiabatic energies will be labeled as $\ket{\epsilon_{0}^{F}}$, $\ket{\epsilon_{1}^{F}}$ and $\ket{\epsilon_{2}^{F}}$. 

To propagate the laser driven electron-nuclear system, we integrate the time-dependent Schr\"{o}dinger equation (TDSE) with the total Hamiltonian $\hat{H}_\text{tot}(\vec{R},t)=-\frac{\hbar^2}{2M}\frac{\partial^2}{\partial R^2}+\hat{H}(\vec{R},t)$ using a Tr\"{o}tter decomposition\cite{kosloff1988time}. 
The nuclear mass is chosen to be $M=1000~\text{a.u.}$.
We assume the initial nuclear wavepacket is on diabat \ket{2} and of the form 
$\ket{\Psi(x, y, 0)}={\cal N}\exp[-\frac{(x-x_{0})^2}{2\sigma_{x}^2} + ip^{x}_{0}(x-x_{0})-\frac{(y-y_{0})^2}{2\sigma_{y}^2} + ip^{y}_{0}(y-y_{0})] \ket{2}$.
Here, $x_{0}, y_{0}, p_{0}^{x}, p_{0}^{y}$ are the initial positions and momenta along the $x$ and $y$ directions respectively, and $\sigma_{x}=\sigma_{y}=1$ represent the width of the Gaussian along $x$ and $y$ direction.
Note that solving the TDSE with the total Hamiltonian is equivalent to the exact propagation of $i\hbar\frac{\partial}{\partial t}C = {H}_FC$ with a time-independent Floquet Hamiltonian. 

\subsection{Floquet-based Berry curvature}
\begin{figure}[tbhp]
    \centering
    \includegraphics{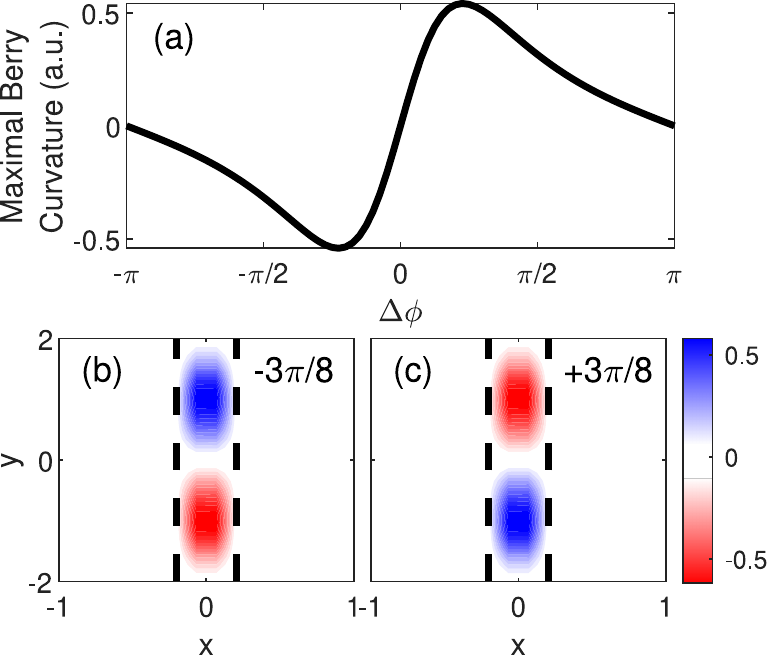}
    \caption{(a) The maximal Berry curvature found in the $y<0$ half-plane for Floquet adiabatic state $\ket{\epsilon_{0}^{F}}$ as a function of $\Delta\phi$. Note that the Berry curvature becomes zero at $\Delta\phi=-\pi,0,\pi$ and reaches the largest magnitude around $\Delta\phi\approx\pm3\pi/8$. The map of the Berry curvature in the nuclear coordinate is plotted for phase differences (b) $\Delta\phi=-3\pi/8$ and (c) $\Delta\phi=+3\pi/8$. Note that for $\Delta\phi=\pm3\pi/8$, the Berry curvatures are the same in magnitude, but opposite in sign, leading to opposite Berry magnetic forces. Note further that these Berry curvature plots have excluded the contribution from the trivial crossing at $x=0$ and $x=\pm \text{atanh}(0.2)$, which are zero in principle (when evaluated exactly).}
    \label{fig: berrycurvature2d}
\end{figure}
Before running dynamics, we analyze the Berry curvature (in the $z$ direction) of the Floquet Hamiltonian as calculated by \eqref{eq:berrycurvature}.
For this 2D model, the Berry curvature is antisymmetric with respect to the $x$ axis, see Fig.~\ref{fig: berrycurvature2d} (b) and (c). In agreement with the theory presented above, the computed Berry curvatures for phase differences $+\Delta\phi$ and $-\Delta\phi$ have the same magnitude but opposite sign. 
For a nuclear wavepacket moving in the $x-y$ plane, equal and opposite Berry curvatures are equivalent to equal and opposite effective magnetic fields, so that we expect the nuclear wavepackets will move in different directions for Hamiltonians specified by phase differences $\pm\Delta\phi$.

Note that the magnitude of the Berry curvature varies dramatically as a function of space. In order to best characterize how one phase difference $\Delta\phi$ determines the overall Berry curvature, in Fig.~\ref{fig: berrycurvature2d}(a), we report the maximum (signed) Berry curvature sampled over the lower half-plane ($y<0$) as a function of $\Delta\phi$. We find the largest difference in the Berry magnetic force when comparing $\Delta\phi\approx3\pi/8$ vs $\Delta\phi\approx-3\pi/8$.
As a sidenote, the Berry curvature is indeed zero when $\Delta\phi=-\pi,0,\pi$---in agreement with the analytic theory discussed in Sec.~\ref{sec:condition}. 

\subsection{Distinct transmission and reflection probabilities as induced by opposite Berry forces}
To observe the consequence of a large Berry magnetic force, we compare the transmission and reflection probabilities of the wavepacket in the presence of the $\omega$ and $2\omega$ CW lasers with phase difference $\Delta\phi=-3\pi/8$ and $\Delta\phi=+3\pi/8$; see Fig~\ref{fig: berrycurvature2d}(a). 
We initialize an incoming wavepacket centered at $(x_{0}, y_{0}) = (-2, -6)$ with the initial momentum $(p_{0}^{x}, p_{0}^{y}) = (6.7, 9)$. 
We choose the initial conditions so that both (i) the wavepacket will pass through the non-zero Berry curvature region (see Figs.~\ref{fig: berrycurvature2d} (b) and (c)) and (ii) the wavepacket momentum will be slow enough so that the asymptotic wavepacket is not sensitive to the initial position of the wavepacket or the initial phase of the $\omega$ and $2\omega$ CW waves (see discussion in Ref~\citenum{zeyu2020arobust}). 

\begin{figure*}
    \centering
    \includegraphics{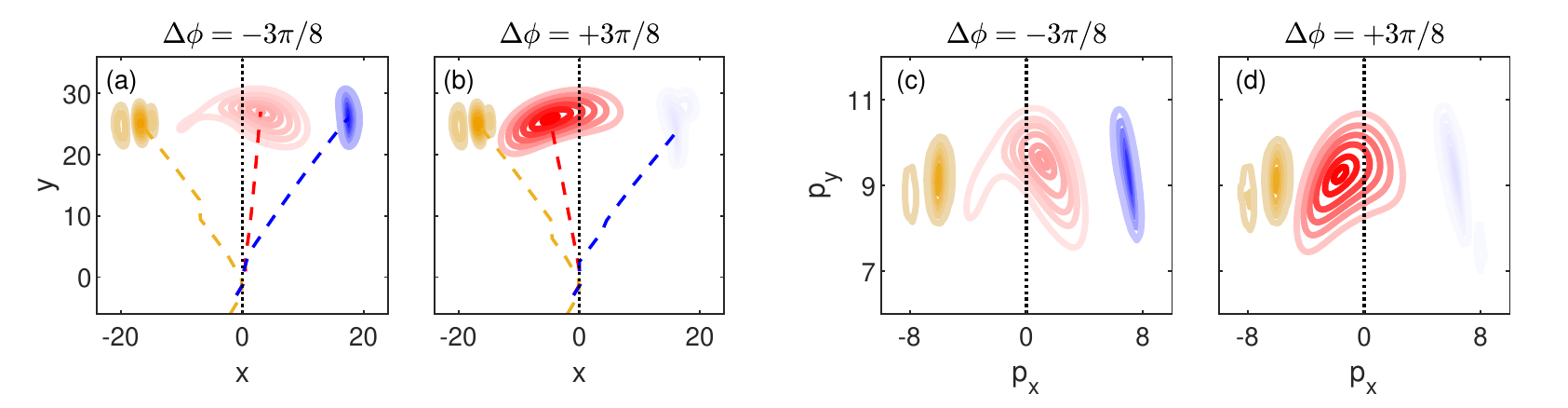}
\caption{Visualization of the transmitting wavepackets in (a)(b) real space and (c)(d) momentum space, corresponding to different diabatic electronic states $\ket{0}$ (Yellow), $\ket{1}$ (Red), and $\ket{2}$ (Blue). For all the panels, the dashed lines represent the trajectories for each states, the dotted black lines represent the $x = 0$ and $p_x=0$ line respectively.
For diabatic states $\ket{0}$, the transmitting wavepackets for $\Delta\phi = -3\pi/8$ and $\Delta\phi = +3\pi/8$ are approximately the same. However, on diabatic state $1$ and $2$, the transmitting wavepackets have significant differences for $\Delta\phi = -3\pi/8$ and $\Delta\phi = +3\pi/8$ in spatial distribution and momentum distribution. 
Moreover, as qualitatively shown by the transparency of the contour colors, the asymptotic probabilities for opposite phase differences are very different. When we combine the probabilities on all states for $x<0$ (or equivalently $p_x$<0) vs $x>0$ ($p_{y}>0$), we find that for $\Delta\phi = -3\pi/8$, Prob($x<0$) = 0.440 and Prob($x>0$) = 0.560; for $\Delta\phi = +3\pi/8$, Prob($x<0$) = 0.939 and Prob($x>0$) = 0.061.
These differences can arise only due to the Berry magnetic force. 
}
\label{fig: wfnks}
\end{figure*}

Finally, let us analyze the nuclear dynamics. As shown in Fig.~\ref{fig: wfnks}, after scattering, the asymptotic wavepackets for the two choices of $\Delta\phi$ are significantly different both in their spatial distributions (Figs.~\ref{fig: wfnks}(a,b)) and in their momentum distributions (Figs.~\ref{fig: wfnks}(c,d)). 
First, we find that, after scattering, the wavepacket on state 1 (red) moves in different directions as a function of $\Delta \phi$ ($x<0$, $p_x<0$ for $\Delta\phi=-3\pi/8$ and $x>0$, $p_x>0$ for $\Delta\phi=+3\pi/8$),which is demonstrable proof that the Berry magnetic force have a strong influence on turning and guiding nuclear dynamics. 
Second, if we focus on the total transmission ($x>0$) and reflection ($x<0$) probabilities (after adding up the contributions from all three electronic states), we find that the bifurcation forward and backward in the x-direction  is very different depending on $\Delta \phi$. In particular, for $\Delta\phi=-3\pi/8$, we find that  $\text{Prob}(x<0)=0.44$ and $\text{Prob}(x>0)=0.56$; vice versa,  for $\Delta\phi=+3\pi/8$, we find  $\text{Prob}(x<0)=0.939$, which implies almost complete reflection. 
And at the same time, the transition probabilities from state 0 to state 1 (red) and 2 (blue) are significantly different as well (see Table~\ref{table1} and the color contours in Fig.~\ref{fig: wfnks}). 
Altogether, these significant differences suggest that the Berry magnetic force effect should be able to promote or suppress a chemical reaction, which should indeed be easy to observe experimentally. 
\begin{table}
\begin{center}
\begin{tabular}{ lcc }
    & $\Delta\phi=-3\pi/8$ & $\Delta\phi=+3\pi/8$ \\
 \hline
 $\text{Prob}(0\rightarrow0)$ & $0.440$ & $0.441$ \\ 
 $\text{Prob}(0\rightarrow1)$ & $0.355$ & $0.061$ \\ 
 $\text{Prob}(0\rightarrow2)$ & $0.205$ & $0.498$ \\ 
 \hline
\end{tabular}
\caption{\label{table1} The transition probabilities for $\Delta\phi=\pm3\pi/8$ as calculated in Fig.~\ref{fig: wfnks}.}
\end{center}
\end{table}

\section{Discussion\label{sec:Discussion}}
We have isolated one class of Hamiltonians whose dynamics clearly demonstrate large effects as caused by the presence of light-induced Berry forces. In order to realize such a Hamiltonian (or a similar Hamiltonian) within a realistic spectroscopic experiment, there are three major questions that we must now address. First, based on the theory in Sec.~\ref{sec:condition}, the experiment must be carried out with molecules or materials that have three bright transitions, i.e. two absorption bands at frequencies $\omega$ and $2\omega$ and one interband transition. What molecules or materials should we choose to satisfy such a requirement?
Second, the intensity of the laser source must be strong enough such that the transition dipolar coupling of the system leads to transitions between diabatic dressed states; how much power must the laser produce in practice?
Third, the experimental measurements must be sensitive to the nuclear dynamics in order to exhibit different signals induced by Berry magnetic force effects; what observables should be measured?
We will now address these practical questions in detail.

\subsection{Molecules and Materials}
As far as target molecules and materials, we can envision several possible candidates for the experiment proposed above:
\begin{enumerate}
    \item Hybrid metal nanostructures: 
    Cyltrimethylammonium bromide (CTAB) is widely used as a surfactant ligand in metallic nanoparticle (NP) synthesis and fortuitously has vibrational bands at $1500~\text{cm}^{-1}$ and $3000~\text{cm}^{-1}$\cite{guivar2015preparation,ding2017plasmon}. Moreover for gold nanoparticles capped with CTAB (CTAB@AuNPs), we can expect very strong absorption for all transitions due to the coupling to plasmons. As such, illuminating a CTAB-coated hybrid nanostructure with IR lasers is one possibility for realizing the experiment above.
    
    \item Photosynthetic complexes: 
    As another example of bright molecules with a fortuitous energy spacing, many light-absorbing components within a photosynthetic bacteria, such as bacteriochlorophyll (BChl) and bacteriophytochromes, have separate absorption bands around $400~\text{nm}$ (Soret band) and $800~\text{nm}$ (the $Q_y$ band)\cite{oren2011characterization, lenngren2018coordination}. In a heterogeneous environment, without symmetry, there is no reason to expect that the Soret band to $Q_{y}$ band should be forbidden.
    \item Quantum dot(QD)--molecule complexes: 
    Lastly, rather than relying on a fortuitous alignment of energies, another approach for generating bright transitions at frequencies $\omega$ and $2\omega$ is to match an adsorbate with a quantum dot of the optimal radius. After all, the electronic properties of a colloidal QD, especially the exciton energy and the transition dipole moment, can be tuned by changing the core size and capping ligands. As an example, oleate-capped colloidal PbSe QDs with a diameter of $6.5~\text{nm}$ have an exciton energy around $5500-6000~\text{cm}^{-1}$ while the oleate ligand itself has a vibrational mode that absorbs strongly at $2900~\text{cm}^{-1}$ \cite{abelson2020collective}. Thus, matching a ligand with a tunable QD is another attractive approach to generating three bright transitions: two at frequency $\omega$ and one at frequency $2\omega$.
\end{enumerate}

\subsection{Laser Source Intensity}
We now turn our attention to the intensity of the laser source as required to observe a Berry force effect in the proposed experiment.
As shown in Fig~\ref{fig: wfnks}, when the initial diabatic state is 0, the dynamical difference of the nuclear wavepackets between the $\Delta\phi=\pm3\pi/8$ cases is most significant on diabats 1 and 2.
Therefore, in addition to having non-zero Berry curvature,
another key requirement for observing Berry force is that the light-induced diabatic coupling (characterized by $D$ in \eqref{eq:diabaticcoupling}) must be strong enough to induce meaningful jumps between diabatic dressed states---for the model Hamiltonian above, we estimate that a lower bound $D>\sqrt{p^x_0A/2\pi M}\approx10^{-3}~\text{a.u.} \approx 0.02~\text{eV}$ will yield a reasonable diabatic transition probability according to the Landau--Zener formula (and a smaller coupling may work as well). 
More generally, given thermal motion at room temperature, a diabatic coupling on the order of $0.02~\text{eV}$ should be quite sufficient for ensuring transitions between diabatic states.
For instance, within the Marcus model of electron transfer, the key Massey parameter that dictates the probability of a transition between diabats is $W\approx 2\pi H_{ab}^2/\hbar\Omega\sqrt{E_R k_B T}$ where $H_{ab}$ is the diabatic coupling, $\Omega$ is the nuclear frequency of the reaction coordinate, and $E_R$ is the reorganization energy\cite{massey1949collisions,tully2012perspective}. 
At room temperature $k_B T=0.025~\text{eV}$, a typical nuclear frequency might be $\hbar\Omega\approx0.01~\text{eV}$ and a typical reorganization energy is about $E_R\approx1.0~\text{eV}$. 
Therefore, one might estimate $H_{ab}>\sqrt{0.01\sqrt{0.025}/2\pi}\approx0.016~\text{eV}$ should imply $W>1$ and a large probability to switch between diabats.

Now, for most photoexcitation experiments, the light-induced diabatic coupling is given by $D=\mu E_L$, where ${\mu}$ is the transition dipole moment, ${E}_L$ is the peak electric field strength, and the average laser power is $P_L=\alpha_L\frac{c\epsilon_0}{2}|\frac{D}{\mu}|^2$ where $\alpha_L$ is the focus area. 
Given that the transition dipole moment of the above candidates ranges from $\mu\approx4.5~\text{Debye}$ (Bchl molecules\cite{knox2003dipole,oviedo2011transition}) to $\mu\approx1000~\text{Debye}$ (colloidal QD\cite{sabaeian2014investigation}) and assuming that the laser focus area is $0.003~\text{mm}^2$ (with a beam waist radius of $30~\mu m$), we can then estimate the necessary laser power that one would need to see a reasonable Berry force effect as somewhere between $1.0$ and $10^{4}$ Watts (see Table~\ref{table2}).
Note that these requirements are somewhat different from the conditions which have historically been applied to create LICIs in gas phase photodissociation experiments; for those experiments, one usually applies a laser pulse having a duration around $10-100~\text{fs}$ and a peak intensity of the order of $10^{12}~\text{W/cm}^2$\cite{corrales2014control,halasz2015direct,kim2012control}). By contrast, for the model proposed above with Floquet theory, the experiments will require a CW laser field (or perhaps a long plateau pulse on the order of ns) with much lower power. These requirements should be realizable given today's laser sources. 
\begin{table}
\begin{center}
\begin{tabular}{ clr } 
    parameters & & values \\
 \hline
 $\mu$&[$\text{Debye}$] & $4.5-1000$ \\ 
 $E_L$&[$\text{V/m}$] & $1.1\times10^8-5.1\times10^5$ \\ 
 $P_L$&[$\text{W}$] & $5.1\times10^{4}-1.0\times10^{0}$ \\ 
 \hline
\end{tabular}
\caption{\label{table2}An estimate of the electric field strength and laser power necessary for observing a reasonably strong Berry force effect given a diabatic coupling between light-dressed Floquet states to be $0.02~\text{eV}$.
A smaller diabatic coupling (i.e. a lower electric field strength and laser power) would likely lead to a detectable Berry magnetic force, but the effect might not be strong. }
\end{center}
\end{table}

\subsection{Physical Observables}
As far as experimental measurements are concerned, we must emphasize that any experimental difference between $H_F$ and $H_F^*$ must reflect a Berry force effect. Thus, one can imagine several different experimental approaches for isolating Berry phase in practice.
\begin{enumerate}
    \item Velocity map imaging (VMI): 
    For a gas phase photodissociation reaction, VMI is one technique for quantifying the kinetic energy distribution of the nuclear fragments that are generated\cite{corrales2014control}. Thus, if a given photodissociation channel can be strongly activated by exposure to two frequencies $\omega$ and $2\omega$ within the lifetime of a molecular beam experiment, VMI should be able to directly quantify fragment momenta as a function of $\Delta\phi$ and in so doing isolate a Berry curvature effect.
    \item Fluorescence emission spectrum: 
    Next, steady-state fluorescence spectra do reflect some degree of excited state dynamics. After all, when exposed to continuous illumination, a molecule or material can go through several transformations before emitting a photon. In this regard, if a molecule or material relaxes differently depending on the Berry curvature, one should expect to measure different emission spectra. In particular, very often excited state dynamics can be probed by measuring the anisotropy of the emission spectra, and this represents another experimental measurement for probing Berry phase effects.
    \item Photo-induced current measurement: 
    When a QD is placed in a nanojunction in the presence of a light field, it is well known that the shape of the current-voltage (I-V) curve is sensitive to the illumination (usually using a single-frequency laser that leads to photo-assisted electron tunneling processes\cite{meyer2007photon,braakman2013photon}. Although the experiments are very difficult, in principle, one can imagine that the photon-induced current can pass through a QD-ligand nanojunction with three bright transitions in the presence of two light fields of frequencies $\omega$ and $2\omega$. Does the Berry magnetic force lead to a strong modified I-V curve?
\end{enumerate}


\section{Conclusions}
In conclusion, physical chemists today have the necessary laser power and non-linear optics such that they should be able to determine whether or not Berry force can have a meaningful effect on chemical dynamics. Here we have proposed the simplest set of experiments to make such a determination.
These experiments require a material with three bright transitions $\ket{0} \xrightarrow[]{\omega} \ket{1}$, $\ket{1} \xrightarrow[]{\omega} \ket{2}$, $\ket{0} \xrightarrow[]{2\omega} \ket{2}$ as well as two strong CW at frequencies $\omega$ and $2\omega$. In the presence of these two laser fields, the key control parameter is the difference in phase $\Delta\phi$. If identical experiments are carried out for $\Delta\phi$ vs $-\Delta\phi$, any detected difference can and must be attributed to Berry force alone. If chemical spectroscopists can indeed isolate such differences and connect to the Berry theory of geometric phase,
such a connection will not only help merging chemical physics and quantum optics, it may also lead to a new understanding of quantum control and photo-chemical catalysis.

\section*{acknowlegment}
{This material is based upon work supported by the U.S. Department of Energy, Office of Science, Office of Basic Energy Sciences under Award Number DE-SC0019397. This research also used resources of the National Energy Research Scientific Computing Center (NERSC), a U.S. Department of Energy Office of Science User Facility operated under Contract No. DE-AC02-05CH11231. We thank Abraham Nitzan, Jessica Anna, Qi Ou, and Shaojie Liu for very helpful discussions.}


\bibliography{reference}

\end{document}